\documentclass[12pt]{article}

\usepackage{amsmath}

\begin{document}

\begin{center}
{\bf     V.M. Red'kov\footnote{E-mail:  redkov@dragon.bas-net.by
}, N.G. Tokarevskaya, V.V. Kisel \\[2mm]
On  the   theory of a scalar particle with  electromagnetic
polarizability\\
in Coulomb and Dirac  monopole  fields\\[2mm]}
 Institute of Physics of
 National   Academy of Sciences of Belarus
  \\
  Belarus State Pedagogical University of M.Tank

\end{center}

\begin{quotation}

15-component  matrix and tetrad-based description of a  a scalar
particle with two electro\-magnetic characteristics -- charge $e$
and  polarizability $\sigma$,  is elaborated in  presence of
external Coulomb field. With the use of Wigner's $D$-functions
technics, in the  basis of diagonal spherical tetrad, the
separation of variables  in the generalized wave equation is done,
and  a system of 15 radial equations is given.  It is shown that
all the radial system is  reduced to a generalized Klein-Fock
radial equation with an additional term of the form $\sigma (e^{4}
/ M^{2} r^{4})$. In the framework of the analogous approach a
scalar particle with charge $e$ and polarizability $\sigma$ is
investigated in presence of the field of a magnetic charge $g$.
The separation of variables is done. Again all the radial system
is reduced  to a single differential equation of second order with
an additional term of the form  $  \sigma \;( e^{2} g^{2} / M^{2}
\; r^{4}) $ . This means that because of the known peculiar
properties of the potential  $r^{-4}$ (an absorbing center), the
monopole influence on the scalar particle with
$\sigma$-characteristics is much more noticeable in the radial
equation than in   the case of usual scalar particle with a charge
only.

\end{quotation}



\section{Introduction}

\hspace{5mm}
This work continues study of the 15-component theory of a charged scalar particle with
additional electromagnetic characteristic -- polarizability [1-10] now turning to
behavior of such a particle  in Coulomb and Dirac magnetic
monopole  fields.

Initial  system of generalized equations for a free scalar
particle of the  mass $m= -iMc / \hbar$ has the  form
\begin{eqnarray}
 \lambda_{1} \; \partial^{a} \; \Phi_{1a} + \lambda_{2} \; \partial^{a} \;
 \Phi_{2a}  - m \;\Phi = 0 \;, \qquad \qquad
\nonumber
\\
 \lambda^{*}_{1}\; \partial_{a} \; \Phi + \lambda_{3}\;  \partial^{b}
\; \Phi_{ba} -  m \; \Phi_{1 a}  = 0\; , \qquad \qquad
\nonumber
\\
 - \lambda^{*}_{2}\; \partial_{a}\; \Phi +
\lambda_{4}\;  \partial^{b} \; \Phi_{ba} -  m \; \Phi_{2 a} = 0 \;, \qquad \qquad
\nonumber
\\
 \pm \;
\lambda^{*}_{3} \; (\partial_{a}\Phi_{1 b} - \partial_{b}\Phi_{1 a}) \mp
\; \lambda^{*}_{4} \; (\partial_{a}\Phi_{2 b} - \partial_{b}\Phi_{2
a}) -  m \; \Phi_{ab}=0 \;. \qquad \qquad
\label{1.1}
\end{eqnarray}

\noindent
Here  $\Phi$ stands for a scalar, $\Phi_{1a}$ and $\Phi_{2a}$ represent two  4-vectors, and
$\Phi_{ab}$ is anti-symmetrical tensor. Latin indices take on the
values $0$, $1$, $2$, $3$. In the  flat Minkowski space the
metric tensor $(g_{ab})=\mbox{diag} (+1, -1, -1, -1)$ is used.
Field functions  $\Phi$, $\Phi_{1a}$, $\Phi_{2a}$, $\Phi_{ab}$ are
assumed to be complex-valued; therefore they describe a charged
scalar particle.

One should take attention to the following fact: because of
complex-valued parameters $\lambda_{1}, ... \lambda_{4}$
separating in eqs. (\ref{1.1}) the corresponding theory for scalar
particle without electric charge will require  special treatment
-- this task will be solved bellow. From previous works [1,2] it
is known that four  parameters $\lambda_{i}
$
can be chosen differently -- it is essential only that they obey
two  conditions
\begin{eqnarray}
\lambda_{1} \; \lambda^{*}_{1} -  \lambda_{2} \; \lambda^{*}_{2} = 1 \;,
 \qquad
\lambda_{3} \; \lambda^{*}_{3} - \lambda_{4} \; \lambda^{*}_{4} = 0 \;  .
\label{1.2}
\end{eqnarray}

\noindent  The above equations  (\ref{1.1}) can be rewritten in a matrix
form (15-component vector-column $\Psi=(\Phi, \;\Phi_{1a},\;\Phi_{2a},\;\Phi_{ab})$)
\begin{eqnarray}
(\; \Gamma^{a}\partial_{a} - m \;  ) \;  \Psi = 0 \;  , \qquad \qquad
\nonumber
\\
%\qquad \qquad \label{1.3} \\
\Gamma^{a}= \left | \begin{array}{cccc}
 0                         &\lambda_{1}G^{a}               & \lambda_{2}G^{a}             & 0 \\
 \lambda^{*}_{1}\Delta^{a} &0                              & 0                            &\lambda_{3}K^{a} \\
-\lambda^{*}_{2}\Delta^{a} &0                              & 0
&\lambda_{4}K^{a} \\
  0                        &\pm \lambda_{3}^{*}\Lambda^{a} &\mp\lambda_{4}^{*}\Lambda^{a} & 0
\end{array}\right | ,
\label{1.4}
\end{eqnarray}

\noindent where four basic matrices  $\Gamma^{a}$ are expressed in
terms of blocks: $G^{a},\; K^{a},\;\Delta^{a},\;\Lambda^{a}$
 of dimension
$1\times4$; $4\times1$; $4\times6$; $6\times4$  respectively:
\begin{eqnarray}
(G^{a})_{(0)}^{\;\;\;\;k} = g^{ak}\; ,  \; \qquad
(\Delta^{a})_{\;\;\;n}^{(0)} = \delta_{n}^{a} \; ,  \qquad
\nonumber
\\
(K_{a})_{n}^{\;\;\;kl}  =   g^{ak} \; \delta_{n}^{l}  - g^{al} \; \delta_{n}^{k}\;,  \qquad
\nonumber
\\
 (\Lambda^{a})_{\;\;nb}^{k} = \delta^{ak}_{nb} =
\delta^{a}_{n}  \; \delta^{k}_{b}  -
               \delta^{k}_{n} \; \delta^{a}_{b} \;     . \qquad
\label{1.5}
\end{eqnarray}

One  can perform, special linear  transformation
over the wave  function $ \Psi'=S \Psi $,  such that a new
equation for  $\Psi'$ does not contain any complex-valued
quantities. The transformation with required properties can be
factorized as follows: $\Psi ' = S \Psi = S_{2} S_{1} \Psi$, where
\begin{eqnarray}
\Psi' = \left | \begin{array}{l}
        \Phi  \\   \Phi_{a}  \\  C_{a}  \\C_{ab}
\end{array}\right | ,\;
S_{1}= \left | \begin{array}{cccc}
       1  &0               &0               &0\\
       0  &\lambda_{1}     &\lambda_{2}     &0\\
       0  &\lambda_{3}^{*} &-\lambda_{4}^{*} &0\\
       0  &0               &0               &I
\end{array}\right |,
\nonumber
\\
S_{2}= \left | \begin{array}{cccc}
       d^{*}  &0               &0               &0\\
       0      &d^{*}           &0               &0\\
       0      &0               &I               &0\\
       0      &0               &0               &I
\end{array}\right |,\;\;
d = \lambda_{1} \lambda_{3} + \lambda_{2} \lambda_{4} \; ,
\label{1.6}
\end{eqnarray}

\noindent here separate components of  $\Psi '$ are designated by
$\Phi , \Phi_{a},$ $ C_{a}, C_{ab} $. In terms of these
constituents eqs. (\ref{1.1}) take the form
\begin{eqnarray}
\partial^{a} \; \Phi_{a}-m \; \Phi=0 \; ,\qquad
\nonumber
\\
\partial_{a} \; \Phi + \sigma \; \partial^{b} \; C_{ba} - m \; \Phi_{a}=0 \; , \qquad
\nonumber
\\
\partial_{a} \; \Phi - m \; C_{a} = 0 \; ,\qquad
\nonumber
\\
\pm \; (\partial_{b} \; C_{a} - \partial_{a} \; C_{b}) - m \; C_{ba} = 0 \;  . \qquad
\label{1.7}
\end{eqnarray}

\noindent On should take notice that now all dependence on
$\lambda_{i}$-parameters in  the wave  equations (\ref{1.7}) is realized
through only real-valued characteristic $\sigma = d d^{*}$.
Again eqs.  (\ref{1.7}) can be  expressed as one matrix relationship
\begin{eqnarray}
( \; \Gamma'^{a} \; \partial_{a}  - m \; ) \; \Psi' = 0 \; ,\qquad
\nonumber
\\
\Gamma'^{a}= \left |
\begin{array}{cccc}
 0                         &G^{a}   & 0            & 0           \\
 \Delta^{a}                &0       & 0            &\sigma K^{a} \\
 \Delta^{a}                &0       & 0            &0            \\
  0                        &0       &\pm\Lambda^{a} & 0
\end{array}\right | .
\label{1.8}
\end{eqnarray}

\noindent
The favored feature   of this basis is that
now $C$-operation is reduced to only complex conjugation  $\Psi^{'(c)}= C'
(\Psi ')^{*} = \Psi^{'*}$.

Now we should dwell upon some peculiarities of that generalized theory of a scalar particle in
presence of  electromagnetic fields.
The presence of such external fields $A_{a}(x)$ can be taken into account through
extension of the  derivative:
$
\partial_{a}  \; \Longrightarrow  \; D_{a}= \partial_{a} - i {e \over  c\hbar} \; A_{a}, \;\;
$,
where  $e$ stands for a  particle charge. Then eqs. (\ref{1.7})
will take on  the form
\begin{eqnarray}
D^{a} \; \Phi_{a} = m \;\Phi\; , \nonumber
\\
D^{a} \; \Phi + \sigma \;  D^{b}  C_{ba} =   m \; \Phi_{a} \; ,
\nonumber
\\
D_{a} \; \Phi = m \; C_{a} \; ,
\nonumber
\\
\pm \; (D_{b}  \; C_{a} - D_{a} \; C_{b} ) =  m \;  C_{ba} \; .
\label{1.11}
\end{eqnarray}

\noindent Between components of the  wave function one can pick up the main ones $\Phi$,
$\Phi_{a}$  and two auxiliary  $C_{a}$, $C_{ab}$. Excluding
both auxiliary ones from equation:
\begin{eqnarray}
C_{a} = \frac{1}{m}\partial_{a}\Phi \; , \qquad
C_{ab} = \mp \;  { i e \sigma \over m^{2} \hbar c } \; F_{ba} \; \Phi
\label{1.12a}
\end{eqnarray}

\noindent ($F_{ab}=\partial_{a}A_{b}-\partial_{b}A_{a}$ stands for electromagnetic filed
tensor) one  arrives  at  the following equations for  $\Phi$ and  $\Phi_{a}$:
\begin{eqnarray}
D^{a} \; \Phi_{a} = m \; \Phi \;, \qquad
\nonumber
\\
D_{a} \; \Phi \; \mp \; \frac{ie\sigma } {m^{2}\hbar c } \; \; D^{b} \; (F_{ba}\Phi )
= m \; \Phi _{a} \; .
\label{1.12b}
\end{eqnarray}

      Thus,  in the frames of extended theory of a scalar particle, they  describe
a  particle that besides of the electrical charge carries an additional  electromagnetic
characteristic.
In a free particle case this additional characteristic $\sigma$
does not manifest itself in the main equations
(\ref{1.12b}) and auxiliary constituents satisfy conditions
\begin{eqnarray}
C_{a}(x)  = \Phi_{a} (x) \; , \qquad  C_{ab} (x) = 0 \; .
\label{1.12c}
\end{eqnarray}

Now we briefly outline extension of the above theory to general relativity.
In accordance with the known procedure by Tetrode-Weyl-Fock-Ivanenko [11-14]
matrix equation (\ref{1.8}) is to be replaced by
\begin{eqnarray}
 [ \; \Gamma ^{\alpha
}(x)\; ( \partial_{\alpha} \;  +  \; B_{\alpha }(x) ) \; - m \;  ]
\;\Psi  (x)  = 0 \; ,
\label{1.19}
\\
 \Gamma ^{\alpha } =
\Gamma ^{a} e ^{\alpha }_{(a)} \; , \; B_{\alpha } = {1
\over 2}\; J^{ab} e ^{\beta }_{(a)}\nabla _{\alpha }( e_{(b)\beta
}) \; .
\label{1.20}
\end{eqnarray}

\noindent
Where $e^{\alpha }_{(a)}(x) $  is a tetrad a curved space-time with metric tensor
$g_{\alpha \beta }(x)$;
 $J^{ab}$ --- stands for generators of the  given representation $(\Phi  , \Phi_{k}, C_{k},
C_{kl})$
of the Lorentz  group:
$$
J^{ab} = \left | \begin{array}{cccc} 0  &   0  &  0
&  0  \\ 0  &   V^{ab} &  0  &  0  \\ 0  &  0  &  V^{ab} &  0  \\
0  &  0  &  0  &  (V \otimes V )^{ab}
\end{array}  \right | \; .
$$

Because equation (\ref{1.19}) contains a tetrad explicitly, there must exist definite
connection between two different tetrads-based representations, otherwise   eq. (\ref{1.19}) is incorrect.
Such a  symmetry of generally covariant equation can be readily
proved  (all details are omitted here):
equations based on  two different tetrads  related to each other by
$e'=L(x)e$ can be transformed to each other by means of a local
($ 15\times15 $)-gauge matrix
\begin{eqnarray}
\Psi ' (x) = S(x)\; \Psi (x)     \; ,  \;\;
 \left |
\begin{array}{c} \Phi'   \\  \Phi'_{k} \\   C'_{k} \\ C'_{kl}
\end{array} \right |
 =
\left | \begin{array}{cccc} 1 &  0  &  0  &  0  \\ 0 &
L_{k}^{\;\;l} & 0  &  0  \\
0 &  0  &  L_{k}^{\;\;l} & 0  \\
0 & 0  &  0  &  L_{k}^{\;\;m} L_{l}^{\;\;n}
\end{array} \right | \;\;
\left | \begin{array}{l} \Phi  \\ \Phi_{l}  \\    C_{l} \\ C_{mn}
\end{array} \right |                \; .
\label{1.21}
\end{eqnarray}

It can be shown that equation (\ref{1.19}) can be reverted to a tensor form:
\begin{eqnarray}
 \nabla_{\alpha}
\Phi^{\alpha} - m \Phi=0,
\nonumber
\\
\nabla_{\alpha} \Phi + \sigma \;
\nabla^{\beta} C_{\beta \alpha} - m \Phi_{\alpha}=0,
\nonumber
\\
\nabla_{\alpha} \Phi - m C_{a} = 0 ,
\nonumber
\\
\pm \; (\nabla
_{\beta} C_{\alpha} - \nabla_{\alpha} C_{\beta})- m C_{\beta
\alpha}=0 .
\label{1.22}
\end{eqnarray}

\noindent Here $\nabla_{\alpha}$ stands for a generally covariant derivative.
Connection between constitutions of $\Psi$ in (\ref{1.19}) and those in (\ref{1.22}) looks as follows
\begin{eqnarray}
\Phi_{\alpha}  = e_{\alpha}^{(l)} \Phi_{l} \; , \qquad C_{\alpha}
= e_{\alpha}^{(l)} C_{l} \; ,
\qquad
 C_{\alpha \beta }  =
e_{\alpha}^{(m)} e_{\alpha}^{(n)} C_{mn} \; .
%\eqno(1.23)
\nonumber
\end{eqnarray}

It should be added that equations (\ref{1.12a}) and (\ref{1.12b})
will be in effect in Riemannian space as well, with only formal change of
the usual derivatives into covariant ones. Any specific gravitational interaction terms do not
arise here, but in case of any space with non-zero  torsion such a term would be.

\section{ Basic equation and notation}

Let us consider equation (\ref{1.19}) in spherical coordinates of flat
Minkowski space ($x^{\alpha}=(c t, r, \theta, \phi)$)
\begin{eqnarray}
 dS^{2}= c^{2}
dt^{2}-dr^{2}-r^{2}(d\theta^{2}+\sin^{2}{\theta}d\phi^{2}) \;  ,
\nonumber
\\
 \; g_{\alpha\beta}=\left
| \begin{array}{cccc}
                  1  &  0       &  0       &   0 \\
                  0  &  -1      &  0       &   0 \\
                  0  &  0       &  -r^{2}  &  0  \\
                  0  &  0       &  0   &   -r^{2}\sin^{2}{\theta}
                        \end {array}
                \right |
\label{2.1}
\end{eqnarray}

\noindent and in the diagonal spherical tetrad
\begin{eqnarray}
e^{\alpha}_{(0)}=(1, 0, 0, 0) \;, \;\; e^{\alpha}_{(3)}=(0, 1, 0,
0)\; ,
\nonumber
\\
e^{\alpha}_{(1)}=(0, 0, \frac {1}{r}, 0) \;, \;\;
e^{\alpha}_{(2)}=(1, 0, 0, \frac{1}{r \sin \theta})  \; .
\label{2.2}
\end{eqnarray}

With the use of explicit form of the  tetrad and the Ricci coefficients
we get to the following generalized matrices $\Gamma^{\alpha}(x)$
and connection $B_{\alpha}(x)$:
\begin{eqnarray}
\Gamma^{\alpha}(x) =
(\Gamma^{0}, \; \Gamma^{3}, \; \frac{1}{r} \; \Gamma^{1}, \;
\frac{1}{r \sin {\theta}} \; \Gamma^{2}) \; , \;\;\; B_{0}=0 \;,
\nonumber
\\
 B_{r}=0, \;\;  B_{\theta}=J^{31}\;,\; \;\;
B_{\phi}=\sin {\theta}J^{32}+\cos {\theta} J^{12} \; .
%\eqno(2.6)
\nonumber
\end{eqnarray}

\noindent Equation (\ref{1.7}) will take the form
\begin{eqnarray}
\left [ \; \Gamma^{0}
\partial_{0}  +   \Gamma^{3}\partial_{r} + \frac{\Gamma^{1}
J^{31} + \Gamma^{2} J^{32} } {r}  +  { \Sigma_{\theta,\phi }\over r}
  -  m  \;\right   ] \;  \Psi  = 0
\label{2.7}
\end{eqnarray}

\noindent where $\Sigma_{\theta,\phi}  $  stands for an angular operator
\begin{eqnarray}
\Sigma_{\theta,\phi} = \Gamma^{1} \;
\partial_{\theta} \; + \; \Gamma^{2} \; \frac{\partial_{\phi} +
\cos{\theta}J^{12}}{\sin{\theta}} \; .
\label{2.8}
\end{eqnarray}

In the following we will  need explicit form of all matrices
involved. With  the  use of the  notation
\begin{eqnarray}
\vec{e}_{1}=  (  1, 0,  0) \;, \;
\vec{e} _{2}= (  0, 1,  0) \;,\;
\vec{e}_{3} = (  0, 0,  1) \; ,
\nonumber
\\
\vec{e}^{\;\;t}_{1}=\left | \begin{array}{c}
         1  \\  0  \\     0
         \end {array}  \right | ,   \;
\vec{e}^{\;\;t}_{2} = \left | \begin{array}{c}
         0  \\  1  \\     0
         \end {array} \right |, \;
\vec{e}^{\;\;t}_{3} =\left | \begin{array}{c}
         0  \\  0  \\ 1
         \end {array} \right | \; ,
\nonumber
\\
 \tau_{1}= \left | \begin{array}{rrr}
      0  &  0  &     0\\
      0  &  0  &    -1\\
      0  &  1  &     0
         \end {array} \right | , \;
\tau_{2}= \left | \begin{array}{rrr}
      0  &  0  &     1\\
      0  &  0  &     0\\
     -1  &  0  &     0
         \end {array} \right |,\;
         \tau_{3}= \left | \begin{array}{rrr}
      0  &  -1 &     0\\
      1  &  0  &     0\\
      0  &  0  &     0
         \end {array} \right |
\label{2.11}
\end{eqnarray}

\noindent the matrices $\Gamma^{a}$
are
 \begin{eqnarray}
\Gamma^{0} = \left | \begin{array}{ccccccc}
      0  &  1  &    \vec{0}  &  0       &  \vec{0}       &   \vec{0}  & \vec{0} \\
      1  &  0  &    \vec{0}  &  0       &  \vec{0 }      &   \vec{0}   & \vec{0} \\
      \vec{0}^{\;t}   &  \vec{0}^{\;t}  &    0 &   \vec{0}^{\;t}     &  0       &  -\sigma I  & 0 \\
1         &  0  &    \vec{0}  &  0       &  \vec{0}        &
\vec{0}  & \vec{0} \\ \vec{0}^{\;t}  &  \vec{0}^{\;t}   &    0 &
\vec{0}^{\;t}  &  0       &  0  & 0\\
      \vec{0}^{\;t}    &  \vec{0}^{\;t} &    0 &  \vec{0}^{\;t}  &  \pm \; I       &   0  & 0\\
      \vec{0}^{\;t}  &   \vec{0}^{\;t}  &    0 &  \vec{0}^{\;t}  &  0       &   0  & 0
\end {array}
\right |,
%\eqno(2.13)
\nonumber
\end{eqnarray}
\begin{eqnarray}
\Gamma^{i}   \qquad \qquad \qquad  \qquad
\nonumber
\\
= \left |
\begin{array}{ccccccc}
      0  &  0  &   -\vec{e}_{i} &  0       &  \vec{0}       &  \vec{0}         & \vec{0}      \\
      0  &  0  &    \vec{0}     &  0       &  \vec{0}       &   -\sigma \vec{e}_{i}  & \vec{0}       \\
      \vec{e}_{i}^{\;t}  &  \vec{0}^{\;t}  &    0     &   \vec{0}^{\; t}        &  0       &   0    & -\sigma \;\tau_{i}\\
      0  &  0  &    \vec{0}    &  0       &  \vec{0}       &    \vec{0}  & \vec{0}       \\
      \vec{e}^{\;\;t}_{i}  &  \vec{0}^{\;t}  &    0  &   \vec{0}^{\;t}   &  0       &   0      & 0 \\
      \vec{0}^{\;t}   &  \vec{0}^{\;t}  &    0     &  \mp \vec{e}^{\;t}_{i}  &  0       &   0      & 0       \\
      \vec{0}^{\;t}  &  \vec{0}^{\;t}  &    0     &  \vec{0}^{\;t}       &  \pm \tau_{i}&   0      & 0
\end {array} \right | .
%\eqno(2.14)
\nonumber
\end{eqnarray}

We will need explicit form of generators
\begin{eqnarray}
J^{ab} = \left | \begin{array}{cccc} 0  &  0      &  0       &  0
\\ 0  & V^{ab} &  0       &  0  \\ 0  &  0      &  V^{ab}  &  0
\\ 0  & 0      &  0       & (V\otimes V)^{ab}
\end{array} \right |    \; ,
\nonumber
\end{eqnarray}
where
$$
(V^{ab})_{k}^{\;\; l}= -g^{al}\delta^{b}_{k}+g^{bl}\delta^{a}_{k}
\;,
$$ $$
 (V^{23})_{k}^{\;\; l} = \left | \begin{array}{rrrr} 0  &
0      &  0       &  0  \\ 0  &  0      &  0       &  0  \\ 0  &
0      &  0       &  -1  \\ 0  &  0      &  1       &  0
\end{array}
 \right | =
\left | \begin{array}{cc} 0  &    0  \\ 0  &    \tau_{1}
\end{array}
 \right | \;  ,
%\eqno(2.15)
$$

\noindent  and so  on. Also
\begin{eqnarray}
[(V \otimes V)^{ab}]_{mn}^{\;\;\;\;sp}= \;
(-g^{as} \; \delta^{b}_{m} \; + \; g^{bs} \; \delta_{m}^{a}) \;
\delta_{n}^{p} \;
+ \; \delta^{s}_{m} \; (-g^{ap} \;
\delta^{b}_{n} \; + \; g^{bp} \; \delta_{n}^{a}) \;  \;,
\nonumber
\end{eqnarray}

\noindent
where  index combinations $01 , \;   02, \; 03 , \; 23 ,
\; 31 , \; 12 $ are used;
further
\begin{eqnarray}
[(V \otimes V)^{23}]_{mn}^{\;\;\;\;sp}
=
\left | \begin{array}{cc}
      \tau_{1} & 0        \\
         0     & \tau_{1}
\end {array} \right |  , \;\; \mbox{and so on} \; .
%\eqno(2.16)
\nonumber
\end{eqnarray}

The total momentum operator (the sum of orbital and spin ones) has
in Cartesian coordinates the ordinary form
\begin{eqnarray}
J_{k} = l_{k} + S_{k} \; , \qquad \qquad
\nonumber
\\
S_{1} = i J^{23} \; , \; S_{2} = i J^{31} , \; S_{3} = i J^{12} \; , \qquad
\nonumber
\\
S_{k} =  i \;[ \; 0 \oplus  ( 0 \oplus \tau_{k}) \oplus ( 0 \oplus \tau_{k})
\oplus ( \tau_{k} \oplus \tau_{k} \; ] \;  .
\label{2.17}
\end{eqnarray}

Now we are to  determine the form of $J_{k}$-operators in
spherical tetrad basis. Two tetrads are related to each other by
means of the law (see (\ref{1.21}) )
$$
 e^{'\alpha}_{(a)} = {\partial x^{'\alpha}
\over \partial x^{\beta}} \; L_{a}^{\;\;b}  \;
e_{(b)}^{\;\;\;\beta} \; , \;
$$

\noindent where $e^{'\alpha}_{(a)}$ is the spherical tetrad,
$e_{(b)}^{\;\;\;\beta}$ is the Cartesian one. Here, the Lorentz
matrix is reduced to a pure rotation
\begin{eqnarray}
O_{i}^{\;\;j} = \left | \begin{array}{ccc} \cos \theta \cos \phi  &  \cos
\theta \sin \phi  & - \sin \theta \\ - \sin \phi  &  \cos \phi
&  0  \\ \sin \theta \cos \phi  &  \sin \theta \sin \phi & \cos
\theta
\end{array} \right | .
\label{2.18}
\end{eqnarray}

\noindent Connection between Cartesian wave faction $\Psi$ and spherical
one $\Psi '$  looks as
\begin{eqnarray}
\Psi '(x)=S \; \Psi (x) \; , \qquad \qquad
\nonumber
\\
S (\theta, \phi)  =
[\; 1 \oplus (  1 \oplus O ) \oplus (1 \oplus O)
\oplus ( O  \oplus O \;) \;] \;  .
\nonumber
\end{eqnarray}

\noindent With the use of the matrix $S(\theta, \phi)$ one transforms operators
$$
J_{i}=l_{i} + S_{i} \; , \qquad J'_{a}=SJ_{a}S^{-1} \;.
$$

Taking into account the known definitions
\begin{eqnarray}
l_{1}=i\;(\sin \phi \;\partial_{\theta}\; +
 \; ctg \; \theta \;  \cos \phi \;\partial_{\phi}) \; ,
\nonumber
\\
l_{2}=i\;(-\cos \phi \;\partial_{\theta}\; + \;
ctg  \; \theta \; \sin  \phi \; \partial_{\phi}) \; , \;
l_{3} = - i \; \partial_{\phi} \;  ,
%\eqno(2.21)
\nonumber
\end{eqnarray}

\noindent
and straightforwardly derived relations
\begin{eqnarray}
O \;i \; \partial_{\theta}\;O^{-1}\; =
\;i\;\tau_{2} \;,  \qquad
\nonumber
\\
O\;i \;  \partial_{\phi} \;  O^{-1} \; = \;i \;( \cos \theta
\;\tau_{3}\; -\; \sin \theta \; \tau_{1} )\;
%\eqno(2.22)
\nonumber
\end{eqnarray}

\noindent  for
$l'_{a}=O\;l_{a}\;O^{-1}$ one gets:
\begin{eqnarray}
l'_{1}=l_{1}-i(\cos \theta \;\cos \phi \; \tau_{1}+ \sin \phi \;
\tau_{2}
\nonumber
\\
+\frac{\cos^{2}\theta}{\sin \theta}\cos \phi \;
\tau_{3}) \; ,
\nonumber
\\
l'_{2}= l_{2}  -  i  (    \cos \theta  \sin \phi \;
\tau_{1}  -   \cos \phi \;  \tau_{2}
\nonumber
\\
+
\frac{\cos^{2}\theta}{\sin  \theta} \sin \phi  \;\tau_{3}) \;,
\nonumber
\\
l'_{3} = l_{3} + \; i \;(\sin \phi \;\tau_{1}  -    \;\cos \theta \;\tau_{3}) \;.
%\eqno(2.23a)
\nonumber
\end{eqnarray}

\noindent In the same manner, for  $\tau'_{a}=O\tau_{a}O^{-1}$ we have
\begin{eqnarray}
\tau'_{1}=  \cos \theta
\cos \phi \;  \tau_{1}  - \sin \phi \;  \tau_{2}  +
\sin \theta  \cos \phi \;  \tau_{3} \; ,
\nonumber
\\
\tau'_{2} = \cos \theta  \sin \phi \; \ tau_{1} -  \cos \phi \;  \tau_{2}
+  \sin \theta  \sin \phi \;  \tau_{3}\; ,
\nonumber
\\
\tau'_{3}=-\;\sin \phi \;\tau_{1}+ \;\cos \theta \;\tau_{3}\;.
%\eqno(2.23b)
\nonumber
\end{eqnarray}

\noindent
Thus we arrive at the final formulas
$J'_{a}$:
\begin{eqnarray}
J'_{1}=l_{1}+\frac{\cos
\phi}{\sin \theta}\;S_{3}, \; J'_{2}=l_{2}+\frac{\sin
\phi}{\sin \theta}\;S_{3},
\label{2.24}
\\
 J'_{3}=l_{3} , \;\; \vec{J\; '}^{2}=\;-\;\frac{1}{\sin \theta}
\partial_{\theta}\;\sin \theta \;\partial_{\theta}\;
\nonumber
\\
+\;
\frac{
-\partial^{2}_{\;\phi}\;+\;2\;i\;\partial_{\phi}\;S_{3}\cos
\theta\;+ \;S^{2}_{3}} {\sin^{2}\theta}\;.
\nonumber
\end{eqnarray}

For the following, it will be convenient to have the
$S_{3}$-matrix in diagonal form.
To this end one needs $\tau_{i}$-basis to be cyclic one [15]:
\begin{eqnarray}
 \Psi '' = U \Psi ' , \qquad \qquad \qquad
 \nonumber
 \\
  U = 1 \oplus (1 \oplus U_{3}) \oplus (1 \oplus U_{3}) \oplus (  U_{3} \oplus U_{3}) \; ,
\nonumber
%\label{2.25a}
\end{eqnarray}

\noindent
where
\begin{eqnarray}
U_{3} = \left |
 \begin{array}{ccc}
- 1 /\sqrt{2}  &  i /  \sqrt{2}  &  0  \\ 0  &  0  &  1  \\ 1 /
\sqrt{2}  &  i  / \sqrt{2}  &  0
\end{array} \right | \; ,\;\;
U^{-1} _{3} = U^{+}_{3} = \left |
 \begin{array}{ccc}
- 1 /\sqrt{2}  &  0  & 1 /  \sqrt{2}    \\ -i / \sqrt{2}  &  0  &
-i / \sqrt{2}   \\ 0    &  1    &  0
\end{array} \right | \; .
\nonumber
\end{eqnarray}

\noindent  One easily derives
\begin{eqnarray}
U_{3}  \tau_{1} U_{3}^{-1} = {1
\over \sqrt{2}} \left | \begin{array}{ccc} 0  &  -i   &  0  \\ -i
&  0  &  -i  \\ 0  &  -i  &  0
\end{array} \right | =  \tau'_{1} \; ,
\nonumber
\\
U_{3}  \tau_{2} U_{3}^{-1} = {1 \over \sqrt{2}} \left |
\begin{array}{ccc} 0  &  -1  &  0  \\ 1  & 0  &  -1  \\ 0  &  1  &
0
\end{array} \right | =  \tau'_{2}\; , \;
\nonumber
\\
U_{3}  \tau_{3} U_{3}^{-1} = - i \; \left |
\begin{array}{rrr} +1  &  0  &  0  \\ 0  &  0  &  0   \\ 0  &  0
&  -1
\end{array} \right | = \tau'_{3} \; . \;
%\label{2.25c}
\nonumber
\end{eqnarray}

\noindent
In this cyclic basis, the
$S_{3}$ matrix is diagonal
\begin{eqnarray}
 S_{3}= \mbox{diag} \; (
0;  \;  0,   \; 1,  \; 0 , \; -1;
 \;
        0 ,   \; 1,  \; 0 , \; -1; \;
        1 , \; 0 , \; -1; \;
        1 , \; 0 , \; -1 \;)
\label{spin}
\end{eqnarray}

\noindent
as required. Also we will need explicit "cyclic" \hspace{2mm} vector
$\vec{e}_{i}$ and $\vec{e}_{i} \; ^{t}$:
\begin{eqnarray}
\vec{e}_{1} = (
-{1 \over \sqrt{2}},  0  , {1 \over \sqrt{2}} ) \; , \;\;\;
\vec{e}_{2} = ( -{i \over \sqrt{2}},  0  ,  -{i \over
\sqrt{2}} ) \;,\;\;
\vec{e}_{3} = ( 0 , 1  , 0) ,
\\
\nonumber
\vec{e}^{\;t} _{1} = \left | \begin{array}{c}
 - {1 \over \sqrt{2}} \\  0  \\  {1 \over \sqrt{2}}
\end{array} \right | \; ,\;
\vec{e}^{\;t}_{2} = \left | \begin{array}{c}
  {i \over \sqrt{2}} \\  0  \\  {i \over \sqrt{2}}
\end{array} \right | \; , \qquad
\vec{e}^{\;t}_{3} = \left | \begin{array}{c}
 0  \\  1  \\  0
\end{array} \right | \; .
% \eqno(2.26)
\nonumber
\end{eqnarray}

\section{ Separation of variables,  radial equations}

\hspace{5mm}
Now we are ready to separate the variables in equation
(\ref{2.7}). All matrices are assumed to be referred to chosen spherical and cyclic basis,
but any reminding of this will be omitted in the  designation below.
Let the wave function $\Psi(x)$ be taken in the form
\begin{eqnarray}
\Psi (x) = \{  \; C (x)
, \; C_{0}(x) ,\; \vec{C}(x) ,
\nonumber
\\
\; \Phi_{0} (x) , \; \vec{\Phi}(x) ,
\; \vec{E}(x) , \vec{H}(x) \; \} \; .
\label{3.1a}
\end{eqnarray}

After simple calculation with the  use of block structure of matrices
involved we  get to
 the system
\begin{eqnarray}
\partial_{0}   C_{0}  -  \vec{e}_{3}  \partial_{r} \vec{C}  -
{1 \over r}  ( \vec{e}_{1}  \tau_{2}  -  \vec{e}_{2}
\tau_{1} )  \vec{C}
\nonumber
\\
- {1 \over r}  (
\vec{e}_{1} \partial_{\theta}  +  \vec{e}_{2}
{\partial_{\phi}  +   \tau_{3} \cos \theta \over \sin \theta}  )  \vec{C} = m  C \; ,
\nonumber
\\
\partial_{0}  C  - \sigma \vec{e}_{3} \partial_{r} \vec{E} -
{\sigma \over r}  ( \vec{e}_{1}  \tau_{2}  -
\vec{e}_{2} \tau_{1} )  \vec{E}
\nonumber
\\
-  {\sigma \over r}  (  \vec{e}_{1}  \partial_{\theta}  +
\vec{e}_{2}  {\partial_{\phi}  +   \tau_{3}  \cos \theta \over \sin \theta}  )
 \vec{E} = m  C_{0} \; ,
\nonumber
\\
 -  \sigma  \partial_{0} \vec{E} +
\vec{e}_{3}^{\;t} \partial_{r}\; C \qquad
\nonumber
\\
-  \sigma  \tau_{3}
\partial_{r}  \vec{H}  -  {\sigma  \over r}  (
\tau_{1}  \tau_{2}  -  \tau_{2}  \tau_{1}  )  \vec{H}
\nonumber
\\
 +  {1 \over r}   ( \vec{e}_{1}^{\;t}
\partial_{\theta} + \vec{e}_{2}^{\;t}  {\partial_{\phi}
\over \sin \theta})  C \qquad
\nonumber
\\
 +  {\sigma \over r}  (
\tau_{1}  \partial_{\theta}  +  \tau_{2} {\partial_{\phi} +
\tau_{3} \cos \theta \over \sin \theta})  \vec{H} = m \vec{C} \; ,
\nonumber
\\
\partial_{0}  C = m  \Phi_{0} \; , \;\;\;
\nonumber
\\
\vec{e}_{3}^{\;t}  \partial_{r}  C  +  {1 \over r}
( \vec{e}_{1}^{\;t}  \partial_{\theta} +
\vec{e}_{2}^{\;t}  {\partial_{\phi} \over \sin \theta} )  C
 = m  \vec{\Phi} \; ,
\nonumber
\\
(\pm)  [  \partial_{0} \vec{\Phi}  -  \vec{e}_{3}^{\;t}  \partial_{r}  \Phi_{0}
\nonumber
\\
 -  {1 \over r}   ( \vec{e}_{1}^{\;t}
\partial_{\theta} + \vec{e}_{2}^{\;t}  {\partial_{\phi}
\over \sin \theta})  \Phi_{0}  ]  = m  \vec{E} \; ,
\nonumber
\\
(\pm) [  \tau_{3}  \partial_{r} \vec{\Phi}  +  {1
\over r}  ( \tau_{1}  \tau_{2}  -  \tau_{2} \tau_{1}
)  \vec{\Phi}  + \nonumber
\\
+   {1 \over r}  (  \tau_{1} \partial_{\theta}  +  \tau_{2} {\partial_{\phi} + \tau_{3}
\cos \theta \over \sin \theta}) \vec{\Phi}  ] = m \vec{H} \;  .
\label{3.1b}
\end{eqnarray}

\noindent From where with consideration of
$$
\tau_{1}  \tau_{2}  -
\tau_{2} \tau_{1} = \tau_{3}  \; , \;\; \vec{e}_{1}
\tau_{2}  -  \vec{e}_{2}  \tau_{1} = 2 \vec{e}_{3} \; ,
$$

\noindent we will have
\begin{eqnarray}
\partial_{0}  C_{0} -  \vec{e}_{3}  ( \partial_{r} +
{2 \over r})    \vec{C} \qquad
\nonumber
\\
-  {1 \over r}  (  \vec{e}_{1}
\partial_{\theta}  +  \vec{e}_{2}
{\partial_{\phi}  +   \tau_{3}  \cos \theta \over \sin \theta}  )  \vec{C} = m  C \; ,
\label{3.2a}
\end{eqnarray}
\begin{eqnarray}
\partial_{0}  C  -  \sigma   \vec{e}_{3} ( \partial_{r}  +
{2 \over r}   )  \vec{E} \qquad
\nonumber
\\
-  {\sigma \over r}  (
\vec{e}_{1}  \partial_{\theta}  +  \vec{e}_{2}
{\partial_{\phi}  +   \tau_{3}  \cos \theta \over \sin
\theta}  )  \vec{E} = m  C_{0} \; ,
\nonumber
\\
-  \sigma \partial_{0}  \vec{E}  + \vec{e}_{3}^{\;t} \partial_{r}
C  -  \sigma  \tau_{3}  (  \partial_{r}  +  {1  \over r}  )  \vec{H}  + \qquad
\nonumber
\\
+  {1 \over r}   (\vec{e}_{1}^{\;t}  \partial_{\theta} + \vec{e}_{2}^{\;t}
 {\partial_{\phi} \over \sin \theta})  C \qquad
 \nonumber
\\
-
 {\sigma \over r}
(  \tau_{1}  \partial_{\theta}  +  \tau_{2}
{\partial_{\phi} + \tau_{3} \cos \theta \over \sin \theta} )
\vec{H} = m  \vec{C}\;  ,
\label{3.2b}
\end{eqnarray}
\begin{eqnarray}
\partial_{0} \; C = m \; \Phi_{0} \; , \;\;\;
\nonumber
\\
\vec{e}_{3}^{\;t}  \partial_{r}  C  +  {1 \over r}
( \vec{e}_{1}^{\;t}  \partial_{\theta} +
\vec{e}_{2}^{\;t}  {\partial_{\phi} \over \sin \theta})  C
 = m  \vec{\Phi} \; ,
\label{3.2c}
\end{eqnarray}
\begin{eqnarray}
(\pm)  [ \partial_{0}  \vec{\Phi}  -  \vec{e}_{3}^{\;t} \partial_{r}  \Phi_{0} \qquad
\nonumber
\\
- {1 \over r}   (\vec{e}_{1}^{\;t}  \partial_{\theta} + \vec{e}_{2}^{\;t}
 {\partial_{\phi} \over \sin \theta}\;)  \Phi_{0}  ]  = m \vec{E}  \;,
\nonumber
\\
(\pm) [  \tau_{3}  (  \partial_{r}  +  {1  \over r} )  \vec{\Phi} \qquad
\nonumber
\\
+  {1 \over r} (  \tau_{1}  \partial_{\theta}  +  \tau_{2} {\partial_{\phi} +
 \tau_{3} \cos \theta \over \sin \theta}) \vec{\Phi}  ] = m  \vec{H} \;  .
\label{3.2d}
\end{eqnarray}

Now let us  proceed to constructing eigen-functions of the operators
$\vec{J}\; ^{2}, J_{3}$. At this we will adhere to
the well established procedure [15]. Allowing for the known defining relationships for
Wigner D-functions [15]
\begin{eqnarray}
 -i \;   \partial_{\phi} D_{-m, s} ^{j} = m \; D_{-m,s}^{j}  \; ,
\;\;
[\; -\frac{1}{\sin
{\theta}}\partial_{\theta}\sin{\theta}\partial_{\theta}
\nonumber
\\
+
\frac{m^{2}+2m\sigma  \cos {\theta}+\sigma^{2}} {\sin^{2}{\theta}}
 ]  D_{-m,s}^{j}= j(j+1)D^{j}_{-m,s}  \;  ,
%\eqno(3.4)
\nonumber
\end{eqnarray}

\noindent
we obtain the most general form of $\Psi$-function
being eigen-functions of $\vec{J}^{2}$ and $J_{3}$:
\begin{eqnarray}
C(x)  =
e^{-i\epsilon t} \;  C (r) \;  D_{0} \; ,\qquad
C_{0}(x) =
+e^{-i\epsilon t}   \; C_{0}(r) \;  D_{0}   \; ,
\qquad \Phi_{0}(x) =
e^{-i\epsilon t}\; \Phi_{0} (x) \; D_{0} \; ,
\nonumber
\\
\vec{C} (x) =
e^{-i\epsilon t} \; \left | \begin{array}{l}
 C_{1}(r)  \; D_{-1}  \\
 C_{2}(3) \; D_{0}    \\
 C_{3}(r) \; D_{+1}      \end{array} \right | \; , \qquad
\vec{\Phi}(x)   = e^{-i\epsilon t}\; \left | \begin{array}{l}
\Phi_{1} (r)\; D_{-1}  \\ \Phi_{2}(r) \; D_{0}   \\ \Phi_{3}(r) \;
D_{+1}  \end{array} \right | \; ,
\nonumber
\\
\vec{E} (x)  =
e^{-i\epsilon t} \; \left | \begin{array}{l} E_{1}(r) \;  D_{-1}
\\ E_{2}(r) \;  D_{0}   \\ E_{3}(r) \;  D_{+1}     \end{array}
\right | \; , \qquad
  \vec{H} (x) = e^{-i\epsilon t} \; \left |
\begin{array}{l} H_{1}(r)  \; D_{-1}  \\ H_{2}(r)  \; D_{0}   \\
H_{3}(r)  \; D_{+1}    \end{array} \right |\; ,
\label{3.5a}
\end{eqnarray}

\noindent   where the shorted notation  $ D_{s}= D_{-m,s}^{j}(\phi,\;\theta,\;0) \; ,
\;\; s=0,\;+1,\;-1 \;$ is used.
We will  need several recurrent formulas  [15]:
\begin{eqnarray}
\partial_{\theta} \;  D_{-1} =   (1/2) \; ( \;
a  \; D_{-2} -    \nu  \; D_{0} \; ) \; ,
\nonumber
\\
\frac
{-m+\cos{\theta}}{\sin {\theta}} \; D_{-1} = (1/2) \; ( \; - a \;
D_{-2} - \nu \; D_{0} \; ) \; ,
\nonumber
\\
\partial_{\theta}  \; D_{0} = (1/2) \; ( \;
\nu  \; D_{-1} - \nu  \; D_{+1} \; ) \; ,
\nonumber
\\
\frac {-m}{\sin
{\theta}} \; D_{0} = (1/2) \; (  \; - \nu \; D_{-1} -  \nu  \;
D_{+1} \; ) \; ,
\nonumber
\\
\partial_{\theta} \; D_{+1} =
(1/2) \; ( \; \nu  \; D_{0} - a  \; D_{+2} \; ) \; ,
\nonumber
\\
\frac
{-m-\cos{\theta}}{\sin {\theta}}  \; D_{+1} = (1/2) \; ( \; - \nu
\; D_{0} - \ a  \; D_{+2} \; ) \;  .
%\label{3.5b}
\nonumber
\end{eqnarray}

\noindent where  $$ \nu =  \sqrt{j(j+1)} \; , \;\; a =
\sqrt{(j-1)(j+2)} \; . $$

\noindent
As a first step, one should obtain some intermediate relations:
\begin{eqnarray}
(\; \vec{e}_{1} \; \partial_{\theta} \;  +  \;
\vec{e}_{2} \; { \partial_{\phi} + \tau_{3} \cos \theta \over \sin
\theta } \;  ) \; \vec{C} (x)\;
= e^{-i\epsilon t} \; {\nu \over
\sqrt{2}} \; (C_{1} + C_{3}) \;   D_{0} \; ,
%\eqno(3.6a)
\nonumber
\\
(\; \tau_{1} \; \partial_{\theta} + \tau_{2} \; { \partial_{\phi} +
\tau_{3} \cos \theta \over \sin \theta }\; ) \; \vec{H} (x)
=
e^{-i \epsilon t} \; {i \nu \over \sqrt{2}} \; \left | \begin{array}{c} - H_{2} \; D_{-1} \\ (H_{1} - H_{3} ) \; D_{0} \\
H_{2} D_{+1}  \end{array} \right | \; ,
%\eqno(3.6b)
\nonumber
\\
(\;
\vec{e}^{\;\;t}_{1} \; \partial_{\theta}   \; +  \;
 \; \vec{e}^{\;\;t}_{2} { \partial_{\phi} \over
\sin \theta }\; ) \; C (x)
= -\;  e^{-i \epsilon t }  \;  {\nu
\over \sqrt{2}} \; \left | \begin{array}{c} C \; D_{-1} \\  0  \\
C \; D_{+1}  \end{array} \right | \; .
%\eqno(3.6c)
\nonumber
\end{eqnarray}

\noindent
With the use of these we get to the following radial equations:
\begin{eqnarray}
-i   \epsilon  C_{0}  -   ( {d \over d r}  +  {2
\over r})  C_{2}  -  {\nu  \over r}  ( C_{1}  +
C_{3}  )  =   m  C \; .
\label{3.7a}
\end{eqnarray}
\begin{eqnarray}
-i  \epsilon  C  -
\sigma   ({d \over d r}  +  {2 \over r})   E_{2}
 -  {\nu  \sigma \over r}    ( E_{1}  +   E_{3} )
= m  C_{0}  \;,
\nonumber
\\
i  \epsilon  \sigma  E_{1}  +
\sigma   (  {d \over dr}  +  {1 \over r } ) H_{1}  -
 {\nu \over r}   C  +  {i  \nu  \sigma \over r}
H_{2} = m  C_{1}  \; ,
\nonumber
\\
i  \epsilon \sigma  E_{2}
+  {d \over dr} C  -  {i  \nu  \sigma \over r}
 (H_{1}-H_{3}) = m C_{2} \; ,
\nonumber
\\
i \epsilon  \sigma  E_{3}  -  \sigma (  {d
\over dr}  + {1 \over r}  ) H_{3}
 - {\nu \over r}  C  -  {i  \nu  \sigma \over r}   H_{2} = m  C_{3}  \; .
\label{3.7b}
\end{eqnarray}
\begin{eqnarray}
- i  \epsilon  \; C = m  \Phi _{0} \; , \qquad -
{\nu \over r } \; C = m   \Phi_{1} \; ,
\nonumber
\\
{d \over dr }\; C = m \; \Phi_{2} \; , \qquad
-  {\nu \over r}\;  C = m \; \Phi_{3} \; .
\label{3.7c}
\end{eqnarray}
\begin{eqnarray}
(\pm)
[  - i  \epsilon \; \Phi_{1}  +   {\nu \over r} \;
\Phi_{0}  ] = m \; E_{1} \; ,
\nonumber
\\
(\pm)  [ - i
\epsilon\; \Phi_{2}  -  {d \over dr}\; \Phi_{0} ] = m \;
E_{2} \; ,
\nonumber
\\
(\pm)  [  - i  \epsilon \; \Phi_{3} +
{\nu \over r} \; \Phi_{0}  ] = m \; E_{3} \; ,
\nonumber
\\
(\pm)\; [
-i (\;  {d \over dr } +  {1 \over r} ) \;
  \Phi_{1} - { i   \nu \over r }\;  \Phi_{2}  ] = m \; H_{1}\; ,
\nonumber
\\
(\pm)  {i \nu \over r} \;  ( \Phi_{1}  - \Phi_{3} )
 = m \; H_{2} \; ,
\nonumber
\\
(\pm) [  + i  ( {d \over dr}
 +  {1 \over r}  ) \;  \Phi_{3} +  { i  \nu \over r}\;
\Phi_{2}  ]  = m \; H_{3}\; .
\label{3.7d}
\end{eqnarray}

\noindent
Where $\nu$ stands for $ \sqrt{j(j+1)} / \sqrt{2}$.

\section{Particle with polarizability  in the Coulomb field}

\hspace{5mm}
Taking into account the external Coulomb field can be
done by  means of one formal  changing in the free particle radial system:

$$
A_{a} = ({ze \over r},0,0,0) \; , \;\; \;  \epsilon \;\;
\Longrightarrow \;\; (\epsilon \; +\;  {\alpha \over r} )\;, \;\;
\alpha = ze^{2} \; .
%\eqno(4.1)
$$

\noindent  As a result we have equations:
\begin{eqnarray}
-i   (\epsilon + {\alpha \over r} )
C_{0}  -   ( {d \over d r}  +  {2 \over r}) \; C_{2}
\nonumber
\\
-
 {\nu  \over r} \; ( C_{1}  +   C_{3}  )  =   m \; C \;.
\label{4.2a}
\end{eqnarray}
\begin{eqnarray}
-i \; (\epsilon +
{\alpha \over r })  \; C \; - \;\sigma \;  (\; {d \over d r} \; +
\; {2 \over r} \;) \;  E_{2} \;
\nonumber
\\
- \; {\nu \;  \sigma \over r} \;
(\; E_{1} \; + \;  E_{3}\; ) = m \; C_{0} \; ,
\nonumber
\\
i \;
(\epsilon + {\alpha \over r} )\; \sigma \; E_{1} \; + \;  \sigma\;
( \; {d \over dr} \; + \; {1 \over r }\; )\; H_{1} \;
\nonumber
\\
-  \; {\nu
\over r} \;  C \; + \; {i \; \nu \; \sigma \over r} \;  H_{2} = m
\; C_{1} \; ,
\nonumber
\\
i \; (\epsilon + {\alpha \over r})  \; \sigma
\; E_{2} \; + \; {d \over dr}\; C \;
\nonumber
\\
- \; {i \; \nu \; \sigma
\over r}\;
 (H_{1}\;-\;H_{3}) = m \;C_{2} \; ,
\nonumber
\\
i  ( \epsilon + {\alpha \over r} )\; \sigma \; E_{3}  -
 \sigma ( \; {d \over dr}  + {1 \over r} ) H_{3} \;
\nonumber
\\
 -   {\nu \over r} \;  C  -  {i  \nu  \sigma \over r} \;  H_{2} = m \; C_{3} \; .
\label{4.2b}
\end{eqnarray}
\begin{eqnarray}
- i \; ( \epsilon + {\alpha \over r}) \; C = m \; \Phi _{0}
\; , \;\;  - {\nu \over r } \; C = m \;  \Phi_{1} \; ,
\nonumber
\\
{d \over dr }\; C = m \; \Phi_{2} \; ,
\;\;  -\;  {\nu \over r}\;  C = m \; \Phi_{3} \; .
\label{4.2c}
\end{eqnarray}
\begin{eqnarray}
(\pm)  [  - i  (
\epsilon + { \alpha  \over })  \Phi_{1}  +   {\nu \over r}
\; \Phi_{0}  ] = m \; E_{1} \; ,
\nonumber
\\
(\pm)  [  - i (
\epsilon + { \alpha \over r} )\; \Phi_{2}  -  {d \over dr}\;
\Phi_{0}  ] = m \; E_{2} \; ,
\nonumber
\\
(\pm)  [  - i
(\epsilon + {\alpha \over r} ) \;  \Phi_{3} +   {\nu \over r}
\; \Phi_{0} ] = m \; E_{3} \; ,
\nonumber
\\
(\pm) [ -i (  {d
\over dr } +  {1 \over r}\; ) \;
 \Phi_{1} - { i   \nu \over r }\;  \Phi_{2}  ] = m \; H_{1}\; ,
\nonumber
\\
(\pm)  {i \nu \over r}\;  ( \Phi_{1}  - \Phi_{3} )
  = m \; H_{2} \; ,
\nonumber
\\
(\pm) [  + i  ( {d \over dr}
\; +  {1 \over r} )   \Phi_{3} +  { i  \nu \over r}\;
\Phi_{2}  ]  = m \; H_{3}\; .
\label{4.2d}
\end{eqnarray}

One should note these equations  assume  some additional
conditions on radial functions:
\begin{eqnarray}
\Phi _{3} = + \;  \Phi_{1} \; , \qquad C_{3} = + \; C_{1} \; ,
\nonumber
\\
\qquad E_{3} = + \;E_{1}  \; , \;\; H_{3} = - \; H_{1} \; , \;\;
H_{2} = 0 \; .
\label{4.3}
\end{eqnarray}

Here, one interesting  (from theoretical standpoint) fact should be emphasized. In considering analogous
15-component theory of a vector particle with polarizability
(this work will be published separately)
a discrete operator of P-inversion can be used to simplify the corresponding radial system.
An idea is to rationalize the  above  conditions (\ref{4.3}) in terms  of this symmetry.
So, let us try to diagonalyze simultaneously with
$\vec{J}^{\;2}, J_{3}$
else one, spatial inversion,  operator. In initial Cartesian representation it
is
\begin{eqnarray}
 \hat{\Pi} = 1 \oplus ( 1 \oplus -I ) \oplus (1 \oplus -I )\oplus (-I \oplus +I)\hat{P} \; ,
\qquad
\nonumber
\\
\hat{P} \Psi (\vec{r}) = \Psi (-\vec{r}) \;, \qquad \qquad \qquad
%\eqno(4.4)
\nonumber
\end{eqnarray}

\noindent which after translating
to the spherical tetrad and cyclic basis will look as
\begin{eqnarray}
\hat{\Pi}   =
[\; 1 \oplus (1 \oplus \Pi_{3}) \oplus (1 \oplus \Pi_{3}) \qquad \qquad \qquad
\nonumber
\\
\oplus (\Pi_{3} \oplus -\Pi_{3}) \; ]
\;\hat{P}  ,
\; \Pi_{3}  =    \left | \begin{array}{rrr} 0  &  0  &  -1  \\ 0  &
-1 &   0  \\ -1  &  0  &  0   \end{array} \right | .
\label{4.5}
\end{eqnarray}

The proper values equation $\hat{\Pi} \Psi = P \; \Psi $
 leads to solutions of two types:
\begin{eqnarray}
\underline{P=(-1)^{j+1}} \; , \; \qquad  \qquad \qquad \qquad
\nonumber
\\
 C = 0 \; , \; C_{0} = 0 \; ,
\;\; C_{3} = - C_{1} \; , \; C_{2} = 0 \; ,
\nonumber
\\
\Phi_{0} = 0 \; , \;  \Phi_{3} = - \Phi_{1} \; , \; \Phi_{2} = 0 \; ,
\nonumber
\\
E_{3} = - E_{1} \; , \;  E_{2} = 0\;  ,\;  H_{3}  = H_{1} \; ;
\label{4.7}
\end{eqnarray}
\begin{eqnarray}
\underline{P=(-1)^{j}} , \qquad \qquad  \qquad \qquad \qquad
\nonumber
\\
 C_{3} = + C_{1}
\; , \; \; \Phi_{3} = + \Phi_{1} \; , \;\;
\nonumber
\\
 E_{3} = + E_{1}
\; , \; H_{3} = - H_{1} \; , \; \; H_{2} = 0 \; .
\label{4.8}
\end{eqnarray}

We notice equations (\ref{4.8}) coinciding with (\ref{4.3}), which means that any solution of
radial system provides us with a state of definite parity $P = (-1)^{j}$.
In turn, it is readily verified that another set of restrictions
(\ref{4.8}), being  applied to the above radial system,  will lead to a inconsistent equations.
Such  behavior is just what it must be expected for a spin 0 particle.

Let us proceed with radial system. With the use of (\ref{4.3}) it will take the form
\begin{eqnarray}
-i   (\epsilon +
{\alpha \over r} )  C_{0}  -   ( {d \over d r}  +  {2
\over r})  C_{2}  -  {\nu  \over r}  2  C_{1}   =   m  C
\; ;
\label{4.9a}
\end{eqnarray}
\begin{eqnarray}
 -i  (\epsilon + {\alpha \over r })   C
 - \sigma   ( {d \over d r}  +  {2 \over r} )
E_{2} -  {\nu   \sigma \over r} 2  E_{1}  = m
C_{0}  \;,
\nonumber
\\
 i  (\epsilon + {\alpha \over r} ) \sigma
E_{1}  +   \sigma   (  {d \over dr}+ {1 \over r }
) H_{1}  -   {\nu \over r}   C   = m  C_{1}  \; ,
\nonumber
\\
i  (\epsilon + {\alpha \over r})   \sigma  E_{2}  +  {d
\over dr} C  -  {i  \nu  \sigma \over r}  2H_{1}  = m
C_{2}  \; ;
\label{4.9b}
\end{eqnarray}
\begin{eqnarray}
 - i  ( \epsilon + {\alpha \over
r})  C = m \; \Phi _{0} \; , \qquad - {\nu \over r } \; C = m \;
\Phi_{1} \; ,
\nonumber
\\
 \;\;  {d \over dr }\; C = m \; \Phi_{2} \; ; \qquad \qquad
\label{4.9c}
\end{eqnarray}
\begin{eqnarray}
(\pm)  [  - i  ( \epsilon + { \alpha
\over })  \Phi_{1}  +   {\nu \over r} \; \Phi_{0}  ] = m
\; E_{1} \; ,
\nonumber
\\
(\pm)  [  - i ( \epsilon + { \alpha \over
r} ) \Phi_{2}  -  {d \over dr}\; \Phi_{0}  ] = m \; E_{2}
\; ,
\nonumber
\\
(\pm) [ -i (  {d \over dr } +  {1 \over
r})   \Phi_{1} - { i   \nu \over r }\;   \Phi_{2}  = m
\; H_{1}\; .
\label{4.9d}
\end{eqnarray}

\noindent
 Substituting expressions for  $C_{i}$ from  (\ref{4.9c}) into (\ref{4.9d}),  one finds
\begin{eqnarray}
E_{1} = 0     \; , \;\;  E_{2} = (\pm)  ( - {i \alpha \over
m^{2} \; r^{2} }) \; C \; , \;\; H_{1} = 0 \; .
\label{4.10}
\end{eqnarray}

\noindent
Now, allowing for (\ref{4.10}), one produces
\begin{eqnarray}
m \; C_{0} = -i \;
(\epsilon + {\alpha \over r})  \pm  {i\alpha \sigma \over
m^{2}  r^{2}}  {dC \over dr}
 \; ,
\nonumber
\\
m \; C_{1} =  - {\nu \over r} \; C \; , \;
 m\; C_{2} = {dC
\over dr}  \pm  {\alpha  \sigma \over m^{2}  r^{2}} C\;.
\label{4.11}
\end{eqnarray}

\noindent
And finally, with (\ref{4.11}), from  (\ref{4.9a}), we arrive at an equation
for   $C(r)$ (changing   $m^{2}$  to $-M^{2}$, and $2\nu^{2}$  to  $j(j+1)$):
\begin{eqnarray}
{d^{2} \over dr^{2} } \; C \; + \; {2 \over r}\; {d \over dr}
\; C \;  +  \left  [ (\epsilon + {\alpha \over r } )^{2}  - M^{2} -
 {j(j+1) \over r^{2}}  \pm  {\sigma \alpha^{2} \over
M^{2}  r^{4} } \right  ]  C(r) = 0 \;  .
\label{4.12}
\end{eqnarray}

This  differential equation  describes  a scalar particle with polarizability
in external Coulomb field.

\section{Particle in the external
magnetic monopole field}

\hspace{5mm}
Now we investigate the scalar particle with polarizability in presence of
Dirac magnetic monopole field,  adhering the technics developed in the
previous Section.   Here, the main equation is
$$
[ \; \Gamma ^{\alpha }(x)\; (
\partial_{\alpha} \;  +  \; B_{\alpha } - i {e \over \hbar c}
A_{\alpha} ) \; - {m c  \over \hbar } \;  ] \;\Psi  (x)  = 0 \; ,
%\eqno(5.1a)
$$

\noindent where  4-vector is determined as
(the form of Schwinger monopole potential is used [17], which has been transformed to spherical coordinates
$x^{\alpha} = ( x^{0}, r, \theta, \phi ) \;$)
\begin{eqnarray}
A_{\alpha} = (0,
0,0, \; A_{\phi} = g \; \cos \theta ) \; ,
%\label{5.1b}
\nonumber
\end{eqnarray}

\noindent
$g$  stands for a magnetic charge.
All difference from the task  in previous section consists in one  formal change
\begin{eqnarray}
\partial_{\phi} \; \Longrightarrow \; (\partial_{\phi} \; + \; i\; k \; \cos \theta ) \; , \;\;\;
\mbox{where}
\; \;\; k = {e\;g \over \hbar\; c} \; .
%\label{5.2}
\nonumber
\end{eqnarray}

\noindent
Correspondingly the  wave  equation  leads to
\begin{eqnarray}
\left [\; \Gamma^{0}\;
\partial_{0}\;  + \; \Gamma^{3}\; \partial_{r}\; +\;
\frac{\Gamma^{1} J^{31} \;+ \; \Gamma^{2} J^{32} } {r}    + \;
\frac{1}{r}\; \Sigma^{k}_{\theta,\phi } \; - \; m \; \right ] \; \Psi = 0
\; ,
\label{5.3a}
\end{eqnarray}

\noindent where a modified $\theta, \phi$-dependent operator is defined by
\begin{eqnarray}
\Sigma^{k}_{\theta,\phi} = \Gamma^{1} \; \partial_{\theta}  +
 \Gamma^{2} {  \partial_{\phi} + ( J^{12} + i  k )
\cos \theta   \over \sin \theta } \; .
\label{5.3b}
\end{eqnarray}

Most calculation from previous section is  not to be repeated here, we
can momentarily proceed to evidently modified equations of the type (\ref{3.1b}):
\begin{eqnarray}
\partial_{0}  C_{0}  -  \vec{e}_{3}  ( \partial_{r} +
{2 \over r})   \vec{C}  - \qquad
\label{5.4a}
\\
-  {1 \over r}  (
\vec{e}_{1}  \partial_{\theta}  +  \vec{e}_{2}
{\partial_{\phi}  +   (\tau_{3} + i k)
 \cos \theta \over \sin \theta}  )  \vec{C} = m  C \;  ,
\nonumber
\end{eqnarray}
\begin{eqnarray}
\partial_{0}  C  -  \sigma   \vec{e}_{3} ( \partial_{r}  +
{2 \over r}   )  \vec{E}  - \qquad
\nonumber
\\
-  {\sigma \over r}
( \vec{e}_{1}  \partial_{\theta}  +   \vec{e}_{2}
{\partial_{\phi}  +  ( \tau_{3} +  i k)  \cos \theta
\over \sin \theta}  )  \vec{E} = m  C_{0} \; ,
\nonumber
\\
- \sigma  \partial_{0}  \vec{E}  + \vec{e}_{3}^{\;\;t}
\partial_{r}  C  -  \sigma  \tau_{3}  ( \partial_{r}
+  {1  \over r}  )  \vec{H}  +
\nonumber
\\
+  {1 \over r}
( \vec{e}_{1}^{\;\;t}  \partial_{\theta} +
\vec{e}_{2}^{\;\;t}  {\partial_{\phi}  +  i k  \cos \theta
\over \sin \theta} )  C  -
\nonumber
\\
-  {\sigma \over r}  (
\tau_{1}  \partial_{\theta}  +  \tau_{2} {\partial_{\phi}
+ (\tau_{3} + i k) \cos \theta \over \sin \theta})
\vec{H} = m  \vec{C}  \;,
\label{5.4b}
\end{eqnarray}
\begin{eqnarray}
\partial_{0} \; C = m \; \Phi_{0} \; , \qquad
\vec{e}_{3}^{\;\;t}  \partial_{r}  C \qquad
\nonumber
\\
+  {1 \over r}
( \vec{e}_{1}^{\;\;t}  \partial_{\theta} +
\vec{e}_{2}^{\;\;t}  {\partial_{\phi} +  i k \cos \theta
\over \sin \theta} )  C  = m  \vec{\Phi} \; ,
\label{5.4c}
\end{eqnarray}
\begin{eqnarray}
 (\pm) \; [ \; \partial_{0}\; \vec{\Phi} \; - \;
\vec{e}_{3}^{\;\;t} \; \partial_{r} \; \Phi_{0} \; \qquad
\nonumber
\\
- {1 \over
r}   ( \vec{e}_{1}^{\;\;t}  \partial_{\theta} +
\vec{e}_{2}^{\;\;t}  {\partial_{\phi} + i k \cos \theta
\over \sin \theta})  \Phi_{0}  ]  = m  \vec{E} \; ,
\nonumber
\\
(\pm)\; [ \; \tau_{3} \; ( \; \partial_{r}\; + \; {1  \over r} \;) \; \vec{\Phi} \; \qquad
\nonumber
\\
+
 {1 \over r} (  \tau_{1}  \partial_{\theta}  +  \tau_{2}
{\partial_{\phi} + (\tau_{3}  +  i  k)  \cos \theta \over
\sin \theta})  \vec{\Phi}  ] = m  \vec{H}  \; .
\label{5.4d}
\end{eqnarray}

Now it is the point to choose a suitable substitution for a wave function $\Psi$,
that might
allow for definite symmetry properties of the system in presence of monopole.
There exist three operators satisfying SO(3.R) commutation relations:
\begin{eqnarray}
J^{(k)}_{1}=l_{1}+ \frac{\cos \phi}{\sin \theta}( S_{3}
- k ) \; ,
\nonumber
\\
\; J^{(k)}_{2}=l_{2}+\frac{\sin \phi}{\sin
\theta}(S_{3} - k), \; J^{(k)}_{3}=l_{3} \; .
\label{5.5}
\end{eqnarray}

\noindent
Therefore
$$
J^{(k)}_{3}=l_{3}\; , \qquad
\vec{J}_{(k)}^{\;\;2}=\;-\;\frac{1}{\sin \theta}
\partial_{\theta}\;\sin \theta \;\partial_{\theta}\;+\;
$$
$$
+ \;
\frac{ -\partial^{2}_{\;\phi}\;+\;2\;i\;\partial_{\phi}\; (S_{3}
\; -\; k) \cos \theta\;+ \;(S\; -\; k)^{2}_{3}}
{\sin^{2}\theta}\;.
$$

\noindent
A most general wave function with properties
$$
\vec{J}_{(k)}^{\;\;2}   \Psi^{(k)} _{jm} = j(j+1)
\; \Psi^{(k)} _{jm} \; ,
 \;   J^{(k)}_{3}  \Psi^{(k)}_{jm} = m \; \Psi^{(k)}_{j,m} \; .
%\eqno(5.7a)
$$

\noindent has the structure

\begin{eqnarray}
\Psi (x) = \{  \; C (x) , \; C_{0}(x) ,\; \vec{C}(x) , \qquad
\; \Phi_{0} (x) , \; \vec{\Phi}(x) , \; \vec{E}(x) ,
\vec{H}(x) \; \} \; ,
\nonumber
\\[2mm]
C(x)  = e^{-i\epsilon t} \;  C (r) \;  D_{k} \; , \;
 C_{0}(x) = e^{-i\epsilon t}   \; C_{0}(r) \;
D_{k}   \; , \;
\;\ \Phi_{0}(x) =    e^{-i\epsilon t}\; \Phi_{0}
(x) \; D_{k} \; ,
\nonumber
\\[2mm]
\vec{C} (x) =  e^{-i\epsilon t} \;
\left | \begin{array}{c} C_{1}  \; D_{k-1} \\  C_{2} \;
D_{k}  \\ C_{3} \; D_{k+1}
\end{array} \right | \; , \qquad
\vec{\Phi}(x)   = e^{-i\epsilon t}\; \left | \begin{array}{c}
\Phi_{1} \; D_{k-1} \\ \Phi_{2} \;  D_{k} \\ \Phi_{3} \;
D_{k+1}
\end{array} \right | \; ,
\nonumber
\\
\vec{E} (x)  = e^{-i\epsilon t} \; \left | \begin{array}{c}
E_{1} \;  D_{k-1}  \\ E_{2} \;  D_{k} \\  \; E_{3} \;
D_{k+1}
\end{array} \right | \; ,\qquad
\vec{H} (x) = e^{-i\epsilon t} \; \left | \begin{array}{c}
H_{1} \;  D_{k-1}  \\  H_{2} \;  D_{k}  \\ H_{3} \;
D_{k+1}
\end{array} \right | \; ,
\label{5.10a}
\end{eqnarray}

\noindent   where
$$
D_{s}= D_{-m,s}^{j}(\phi,\;\theta,\;0) \; , \; s = k,\;k+1,\;k-1
\; .
%\eqno(5.10b)
$$
With the  use of recursive formulas  [15]
\begin{eqnarray}
\partial_{\theta }  D_{\kappa -1}  =
(a  D_{\kappa-2} - c  D_{\kappa } )\; , \;
\nonumber
\\
{
-m-(\kappa -1) \cos \theta  \over \sin \theta }  D_{\kappa -1} =
(-a  D_{\kappa -2} - c   D_{\kappa }) \; ,
\nonumber
\\
\partial _{\theta } D_{\kappa }   = \;
(c  D_{\kappa-1} -  d  D_{\kappa +1})\; ,
\nonumber
\\
{- m  -
\kappa  \cos \theta \over \sin \theta }  D_{\kappa } = (-c
D_{\kappa-1} - d   D_{\kappa +1})\; ,
\nonumber
\\
\partial _{\theta }  D_{\kappa +1}  =
(d  D_{\kappa } - b   D_{\kappa +2})\; ,
\nonumber
\\
{-m-(\kappa +1)\cos \theta \over \sin \theta }   D_{\kappa +1}
\; = \; (-d   D_{\kappa } - b D_{\kappa +2})  \; ,
\nonumber
\end{eqnarray}

\noindent where
\begin{eqnarray}
 a = {1 \over 2}  \sqrt{(j + \kappa  -1)(j -
\kappa  + 2)}\;  , \qquad
\nonumber
\\
b = {1 \over 2}  \sqrt{(j - \kappa
-1)(j + \kappa  + 2)} \; , \qquad
\nonumber
\\
c = {1 \over 2}  \sqrt{(j +
\kappa )(j - \kappa + 1)} \; , \qquad
\nonumber
\\
d = {1 \over 2}  \sqrt{(j -
\kappa )(j + \kappa  +1)} \; , \qquad
\nonumber
\end{eqnarray}

\noindent
one finds

\begin{eqnarray}
[ \; \vec{e}_{1}
\partial_{\theta} \;  + \; \vec{e}_{2} {\partial_{\phi} \;+\;
(\tau_{3} \;+\; i\; k )\;  \cos \theta \over \sin \theta }  \; ]
\; \vec{C}
= e^{-i\epsilon t } \; \sqrt{2}  \;(  \; c \; C_{1} \;+
\; d \; C_{3}\; )\; D_{k} \; ,
%\eqno(5.13)
\nonumber
\\
\tau_{1} \;
\partial_{\theta} \; \vec{H} \; +\; \tau_{2}\;  \; {
\partial_{\phi} \; + \; (\tau_{3} \; +  \; i\;k ) \cos \theta
\over \sin \theta }  \;\; \vec{H}
=
 e^{-i \epsilon t}  \;(  i \sqrt{2} ) \;
\left | \begin{array}{c}
                       - c\; H_{2} \; D_{k-1}  \\
                       (c \; H_{1} - d \; H_{3})\; D_{k}  \\
                        + d\;  H_{2} \; D_{k+1}            \end{array} \right | \; ,
\nonumber
\\
(\;  \vec{e}^{\;\;t}_{1} \; \partial_{\theta}
\; +  \;
 \; \vec{e}^{\;\;t}_{2}  \; { \partial_{\phi}  +  i \; k\;  \cos \theta \over
 \sin \theta  } \; ) \;C
 =  e^{-i \epsilon t} \;
 e^{-i \epsilon t}  \; \sqrt{2} \;
\left | \begin{array}{c}
                           - c\; C(r) \; D_{k-1}  \\
                                     0            \\
                           - d \; C(r) \; D_{k+1}      \end{array} \right |.
%\eqno(5.15)
\nonumber
\end{eqnarray}

\noindent
After simple calculation one arrives at
\begin{eqnarray}
-i   \epsilon  C_{0}  -   ( {d \over d r}
+  {2 \over r})  C_{2}  -  {1  \over r}  \sqrt{2}  (
c  C_{1}  +   d     C_{3}  )  =   m  C \; ;
\label{5.16a}
\end{eqnarray}
\begin{eqnarray}
- i  \epsilon  C  -
\sigma   ( {d \over d r}  +  {2 \over r} )   E_{2}
 -  {\sigma \over r} \sqrt{2}     (   c  E_{1}  +
d    E_{3} ) = m  C_{0} \; ,
\nonumber
\\
i  \epsilon  \sigma
 E_{1}  +   \sigma  (  {d \over dr}  +  {1 \over r
} ) H_{1}  -   {\sqrt{2}  c \over r}   C
+  {
\sigma \over r}  (i \sqrt{2}c)  H_{2} = m  C_{1}  \; ,
\nonumber
\\
i  \epsilon  \sigma  E_{2}  +  {d \over dr} C  -
 {\sigma \over r} (i  \sqrt{2})  (c H_{1}-  d
H_{3}) = m C_{2} \;  ,
\nonumber
\\
i   \epsilon  \sigma  E_{3}
-  \sigma (  {d \over dr}  + {1 \over r} ) H_{3}
 -  {\sqrt{2} d  \over r}   C
 -  { \sigma \over r}  (i \sqrt{2}  d)
  H_{2} = m  C_{3} \; ;
\label{5.16b}
\end{eqnarray}
\begin{eqnarray}
- i  \epsilon  C = m  \Phi _{0} \; , \qquad - {
\sqrt{2}  c \over r } \; C = m \;  \Phi_{1} \; ,
\nonumber
\\
 {d \over dr }\; C = m  \Phi_{2} \; , \qquad
-  {\sqrt{2} d \over r}  C = m \; \Phi_{3} \; ;
\label{5.16c}
\end{eqnarray}
\begin{eqnarray}
(\pm)  [  -
i   \epsilon  \Phi_{1}  +   {\sqrt{2}  c \over r} \;
\Phi_{0}  ] = m  E_{1} \; ,
\nonumber
\\
(\pm)  [  - i
\epsilon  \Phi_{2}  -  {d \over dr}\; \Phi_{0}  ] = m \;
E_{2} \; ,
\nonumber
\\
(\pm)  [  - i  \epsilon   \Phi_{3} +
{\sqrt{2} d  \over r} \; \Phi_{0}  ] = m \; E_{3} \; ,
\nonumber
\\
(\pm) [ -i  (  {d \over dr }  +  {1 \over r} )
\Phi_{1} -  { i   \sqrt{2}  c \over r } \; \Phi_{2}  ] \ = m \; H_{1}\; ,
\nonumber
\\
(\pm) (i \sqrt{2})\;  {1 \over r}   \;(  c  \Phi_{1}
 -  d  \Phi_{3} )   = m \; H_{2} \; ,
\nonumber
\\
(\pm) [  +
 i  ( {d \over dr}  +  {1 \over r} )   \Phi_{3} +
 { i  \sqrt{2} \; d \over r}\; \Phi_{2} ]  = m \; H_{3}\; .
\label{5.16d}
\end{eqnarray}

Our aim is a final single differential equation for  $C(r)$.
As a first step, expressions for  $\Phi_{a}(r)$ from
(\ref{5.16c}) are to be entered equation  (\ref{5.16d}):
\begin{eqnarray}
m^{2}
 E_{1} =  (\pm) [ ( i\epsilon )  ({ 1 \over r})
\sqrt{2} c  C  +  {\sqrt{2} c \over r} (-i
\epsilon)  C  ]  = 0 \; ,
\nonumber
\\
m^{2}  E_{2} = (\pm)
[ -i  \epsilon  {d \over dr} C - {d \over dr}  (-i
\epsilon)  C   ] = 0 \; ,
\nonumber
\\
m E_{3} = (\pm)  [
i\epsilon {\sqrt{2} c \over r}  C + {\sqrt{2} c \over
r}  (-i \epsilon)  C  ] = 0 \; .
\nonumber
\end{eqnarray}

\noindent Further, let us substitute   $\Phi_{a}(r)$ from (\ref{5.16c})  into
 (\ref{5.16d}):
\begin{eqnarray}
m^{2} H_{1} = (\pm ) [
 -i ({d \over dr} + {1 \over r}) (-{1\over r} )
\sqrt{2}  c  C
 + {1 \over r} \; (-i \sqrt{2}c )
 {d \over dr } C  ]  = 0 \; ,
\nonumber
\\
m^{2}  H_{2} =
(\pm)  {1 \over r}   (i \sqrt{2})  [  - c {1 \over
r} \sqrt{2}  c  C +  d  {\sqrt{2} d \over r}   C
 ]
=
\nonumber
\\
= (\pm)  {2i \over r^{2} } (d^{2} -
c^{2} ]\; C  = (\pm)(-i)  {k \over r^{2} } C \;  ,
\nonumber
\\
m^{2} H_{3} = (\pm ) [ - i ({d \over dr}  +  {1\over r})
{1 \over r} \sqrt{2}  d  C
  + {i d  \sqrt{2} \over r }  {d \over dr} C  ] = 0 \; .
\nonumber
\end{eqnarray}

\noindent Therefore
\begin{eqnarray}
E_{1} = 0  \; , \;
E_{2} = 0 \; ,\;  E_{3} = 0 \; , \; H_{1} = 0 \; , \;
\nonumber
\\
H_{2} = (\pm) (-i)  { k \over m^{2} r^{2}} \; C\; , \;
H_{3} = 0 \; , \;\;
\label{5.17}
\end{eqnarray}

\noindent  which being taken into equations (\ref{5.16b})  give for $C_{a}(r)$:
\begin{eqnarray}
m C_{0} = -i  \epsilon  C  \; ,
\;
m C_{1} = -
{\sqrt{2} c \over r  }   C  \pm \sigma  {
\sqrt{2} c  \over r}  {  k \over m^{2} r^{2} }  C  ,
\nonumber
\\
m  C_{2} = {d \over dr}  C  \;  , \;
m C_{3} = -
{\sqrt{2} d  \over r  } C  \mp \sigma  {
\sqrt{2} d  \over r} {  k \over m^{2} r^{2} }  C \; .
\label{5.18}
\end{eqnarray}

\noindent
Finally, with the use of (5.18), we arrive at
$C(r)$:
\begin{eqnarray}
{d^{2} \over dr^{2} }\; C  +
{2 \over r}\; {d \over dr} \; C \;  +
 [
\epsilon^{2} \; + \;  m^{2} \; -
\nonumber
\\
 - \; {2  (c^{2}  + d^{2}) \over r^{2}}  \pm   \sigma \;  { k \;  \over  m^{2} \;
r^{4} } \; 2  (c^{2} \; - \; d^{2})
    ]\; C = 0 \; .
\nonumber
\end{eqnarray}

\noindent From this, taking the formulas
$$
 c^{2} -  d^{2} = {k \over
2} , \;
 c^{2}  +  d^{2}   = { j(j+1) - k^{2} \over 2}  , \;  m^{2} = - M^{2}\;  ,
$$

\noindent  one gets  a differential equation for $C(r)$
\begin{eqnarray}
{d^{2} \over
dr^{2} }\; C  \;+ \; {2 \over r}\; {d \over dr} \; C \;  +
 \; [ \;
\epsilon^{2} \; - \;  M^{2} \;
\nonumber
\\
-
 \; { j(j+1) - k^{2} \over r^{2} }
\;  \mp  \; \sigma \; { k^{2}  \;  \over  M^{2} \; r^{4} } \;  ]\;
C = 0 \;
\label{5.19}
\end{eqnarray}

\noindent
where  the term proportional to  $\sigma$-parameter describes peculiarity supplied
by additional electromagnetic structure of a scalar particle --- polarizability.
It should be specially noted that because of the known singular properties of a
$r^{-4}$ potential in Shr\"{o}dinger problem the
influence of the monopole on a scalar particle with polarizability will be much more
significant than in the case of ordinary scalar particle;
the latter  formally follows from (\ref{5.19}) on making $\sigma =0$ and  vanishing $r^{-4}$ term.

\section{ Particle in presence of  both electric and magnetic
charges}

\hspace{5mm}
Now we consider behavior of a scalar  particle with polarizability in presence both  of a Coulomb
potential and Dirac monopole one.
One does not need to perform much calculation in addition to yet done in previous sections.
It suffices to make one formal change in  radial equation of previous Section:
\begin{eqnarray}
\epsilon \qquad  \Longrightarrow \qquad
(\epsilon + {\alpha \over  r}) \; .
\nonumber
\end{eqnarray}

\noindent
As  result, one gets
\begin{eqnarray}
 -i \;(\epsilon +
{\alpha \over  r}) \; C_{0} \; - \; ( {d \over d r} \; + \; {2
\over r}) \; C_{2} \;
\nonumber
\\
- \; {1  \over r} \; \sqrt{2} \; (\; c\;
C_{1} \; +  \; d \;     C_{3} \; )  =   m \; C \; ;
\label{6.1a}
\end{eqnarray}
\begin{eqnarray}
- i \; (\epsilon+{\alpha \over r})  \; C \; - \;\sigma \;  (\;
{d \over d r} \; + \; {2 \over r} \;) \;  E_{2} \;
\nonumber
\\
 - \; {\sigma
\over r} \;\sqrt{2} \;    (\;   c \; E_{1} \; + \; d \;   E_{3}\;
) = m \; C_{0} \; ,
\nonumber
\\
i \; (\epsilon+{\alpha \over r})  \;
\sigma \; E_{1}  + i\;  \sigma\;  ( \; {d \over dr} \; + \; {1
\over r }\; )\; H_{1} \;
\nonumber
\\
-  \; {\sqrt{2} \; c \over r} \;  C \; +
\; { \sigma \over r} \; (i\; \sqrt{2}\;c) \; H_{2} = m \; C_{1} \; ,
\nonumber
\\
i \; (\epsilon+{\alpha \over r})  \; \sigma \; E_{2} \; +
\; {d \over dr}\; C \;
\nonumber
\\
-
 \; {\sigma \over r}\;
(i \; \sqrt{2}) \; (c\; H_{1}\;- \; d\; H_{3}) = m \;C_{2} \; ,
\nonumber
\\
i \;  (\epsilon+{\alpha \over r})  \; \sigma \; E_{3}  - i\;
\sigma ( \; {d \over dr} \; +\; {1 \over r}\; )\; H_{3} \;
\nonumber
\\
 -  \; {\sqrt{2}\; d  \over r} \;  C \; - \; { \sigma \over r} \; (i\; \sqrt{2} \; d) \;
  H_{2} = m \; C_{3} \; ;
\label{6.1b}
\end{eqnarray}
\begin{eqnarray}
- i  (\epsilon+{\alpha \over r})   C =
m \; \Phi _{0} \; , \; - { \sqrt{2}  c \over r } \; C = m \;
\Phi_{1} \; ,
\nonumber
\\
 {d \over dr }\; C = m \; \Phi_{2} \; ,
\;
-  {\sqrt{2} d \over r}\;  C = m \; \Phi_{3} \; ; \;\;\;\;
\label{6.1c}
\end{eqnarray}
\begin{eqnarray}
(\pm)  [  - i   (\epsilon+{\alpha \over
r})   \Phi_{1}  +   {\sqrt{2}  c \over r} \; \Phi_{0}  ]
= m  E_{1} \; ,
\nonumber
\\
(\pm)  [  - i  (\epsilon+{\alpha
\over r})   \Phi_{2}  -  {d \over dr} \; \Phi_{0}  ] = m
E_{2} \; ,
\nonumber
\\
(\pm)  [  - i  (\epsilon+{\alpha \over r})
 \Phi_{3} +   {\sqrt{2} d  \over r} \; \Phi_{0}  ] = m  E_{3} \; ,
\nonumber
\\
(\pm) [ -i  (  {d \over dr }  +
{1 \over r} )   \Phi_{1} -
 { i   \sqrt{2}  c \over r }\;  \Phi_{2}  ] \ = m  H_{1}\; ,
\nonumber
\\
(\pm) (i \sqrt{2})  {1 \over r}  (  c  \Phi_{1}
 -  d  \Phi_{3} )   = m  H_{2} \; ,
\nonumber
\\
(\pm) [  +
 i  ( {d \over dr}  +  {1 \over r} )   \Phi_{3} +
 { i  \sqrt{2} d \over r}  \; \Phi_{2} ]  = m  H_{3}\; .
\label{6.1d}
\end{eqnarray}

Substituting expressions for
$\Phi_{a}(r)$ from  (\ref{6.1c})  into eqs. (\ref{6.1d}):
\begin{eqnarray}
 m^{2}
\; E_{1} =  (\pm)\; [ \;( i\;(\epsilon+{\alpha \over r})  )\;\;
({ 1 \over r}) \; \sqrt{2}\; c \; C \;
+ \; {\sqrt{2}\; c \over
r}\; (-i \; (\epsilon+{\alpha \over r}) ) \; C \; ]\;  = \;0 \; ,
\nonumber
\\
m^{2}  E_{2} = (\pm) \; [ \; -i \; (\epsilon+{\alpha \over
r})  \; {d \over dr}\; C
- {d \over dr} \; (-i\; (\epsilon+{\alpha
\over r}) ) \; C   \; ] = \pm \;i\; {(-\alpha) \over r^{2}} \;C \; ,
\nonumber
\\
 m\; E_{3} = (\pm) \; [ \; i(\epsilon+{\alpha \over r})  \;
{\sqrt{2}\; d \over r} \; C
+ {\sqrt{2}\; d \over r}\;
(-i\;(\epsilon+{\alpha \over r}) ) \; C \; ] = 0 \; .
\nonumber
\end{eqnarray}

\noindent Expressions for  $\Phi_{a}(r)$ from are to be substituted into
(\ref{6.1d}):
\begin{eqnarray}
m^{2}\; H_{1} = (\pm )\; [ \; -i\; ({d \over
dr} \;+\; {1 \over r})\; (-\;{1\over r} )\;  \sqrt{2}\;  c \; C
+ {1 \over r} \; (-i  \sqrt{2}c )   {d \over dr
}\; C  ]  =  0 \; ,
\nonumber
\\
 m^{2}  H_{2} = (\pm)  {1 \over
r}   (i \sqrt{2})  [  - c {1 \over r}  \sqrt{2}  c
+ \; d  {\sqrt{2}  d \over r}    ) C
=
   \pm({-i eg \over m^{\;2}
r^{\;2}}) C \;,
\nonumber
\\
 m^{2}  H_{3} = \pm  [ - i
({d \over dr}  +  {1\over r}) {1 \over r} \sqrt{2}
d    + {i d \sqrt{2} \over r }  {d \over dr}   ]C  = 0 \;  .
\nonumber
\end{eqnarray}

\noindent Thus, for  $E_{i}(r),H_{i}(r)$
we get to
\begin{eqnarray}
 E_{1} = 0  \; , \;\; E_{2} = (\pm)\; {-i
\;\alpha \over m^{2}\; r^{2}}\; C \; ,\; \; E_{3} = 0 \; ,
\nonumber
\\
H_{1} = 0 \; , \;\;  H_{2} = (\pm)\; {-i \;eg \over m^{2}\;
r^{2}}\;C \;\;
 , \; \; H_{3} = 0 \; .
\label{6.2}
\end{eqnarray}

\noindent  Accounting  (\ref{6.2}), from   (\ref{6.1b})  one obtains
expressions for $C_{a}(r)$:
\begin{eqnarray}
m  C_{0} = -i  (\epsilon+{\alpha \over r})  \; C
- \sigma ({d \over dr}+ {2 \over r})  (\pm){-i \alpha
\over m^{2} r^{2}}  C \; ,
\nonumber
\\
m  C_{1} = - {\sqrt{2} c
\over r  } \; C  \pm \sigma  \; { \sqrt{2} c  \over r}
\; { ( -i e g )\over m^{2} r^{2} } \; C \; ,
\nonumber
\\
 m  C_{2} =
{d \over dr}\;  C   +\sigma i  (\epsilon+{\alpha \over r})
\;(\pm){-i \alpha \over m^{2} r^{2}}\; C \;  ,
\nonumber
\\
 m C_{3} = - {\sqrt{2} d  \over r  } \; C  \pm
\;\sigma  \; { \sqrt{2} d  \over r} \; {  i e g \over m^{2}
r^{2} } \; C  \; .
\nonumber
\end{eqnarray}
\begin{eqnarray}
\label{6.3}
\end{eqnarray}

\noindent And finally, with (\ref{6.3}), from  (\ref{6.1a})
we  arrive at a  differential equation for  $C(r)$:
\begin{eqnarray}
 -i  (\epsilon+{\alpha
\over r})(-i (\epsilon+{\alpha \over r})\;C
\nonumber
\\
-\sigma ({d
\over dr}+ {2 \over r}) (\pm){-i \alpha \over m^{2} r^{2}}\; C)-
\nonumber
\\
 -({d \over dr}+{2 \over
r})(\pm \sigma\;(\epsilon+{\alpha \over r}) {\alpha \over
m^{2} r^{2}}\; C +{d C \over dr})-
\nonumber
\\
 -{\sqrt{2} \over
r} (c(-{\sqrt{2} c \over r}\;C \pm
{i \sqrt{2} c \;\sigma \over r} { (-i e g )\over m^{2} r^{2} } \; C) +
\nonumber
\\
+ d (-{\sqrt{2} d \over
r}\;C \pm  {i \sqrt{2} d \sigma \over r} { (+ i e g
)\over m^{2} r^{2} } \; C)=m^{2}C \;
\nonumber
\end{eqnarray}

\noindent or
\begin{eqnarray}
- (\epsilon+{\alpha \over r})^{2} \;C \pm {\sigma \alpha^{2}
\over m^{2} r^{4}} C - ({d \over dr}+{2 \over r}) {d C
\over dr} +
\nonumber
\\
 + {2 \over r^{2}} (d^{2}+c^{2}) C \pm
{2\sigma eg (d^{2}-c^{2}) \over m^{2}r^{4}}C= m^{2}C \; .
\label{6.4}
\end{eqnarray}

\noindent Eq. (\ref{6.4}), with the use of (also $m^{2} = - M^{2}$)
\begin{eqnarray}
c^{2} -  d^{2} = -{k \over 2} ,
c^{2}  +  d^{2}   = { j(j+1) - k^{2} \over 2}  ,
%\label{6.5}
\nonumber
\end{eqnarray}

\noindent  will take the form
\begin{eqnarray}
{d^{2} \over dr^{2} }\; C  \;  +
\; {2 \over r}\; {d \over dr} \; C \;  +
 \; \left [ \;(\;\epsilon \;+ {\alpha  \over r}) ^{2} \; - \;  M^{2} \; \right.
 \nonumber
 \\
\left.   -
 \; { j(j+1) - k^{2} \over r^{2} }
\;  \pm  \; \sigma \; { e^{\;4}\;-\;k^{2}  \;  \over  M^{2} \;
r^{4} } \;  \right ] \; C = 0 \; .
\label{6.6}
\end{eqnarray}

The structure of the obtained equation permits  by means of  deliberate fitting
values $\alpha$ and $k$
to have  $ \alpha^{2} = k^{2}$, what means that in equation (\ref{6.6}) the singular term  $r^{-4}$
vanishes.

\end{document}